\documentclass[twocolumn,letterpaper]{IEEEAerospaceCLS}  

\usepackage[]{graphicx}    
\usepackage{url}

\newcommand{\ignore}[1]{}  

\begin{document}

\title{A Compact Anomaly Detection Solution for Science Instruments}

\author{%
Alfonso Lagares de Toledo\\ 
School of Aerospace Engineering\\
Georgia Institute of Technology\\
Atlanta, GA 30312\\
alagares@gatech.edu
\and
Christopher E. Carr\\
School of Aerospace Engineering \&\\
School of Earth and Atm. Sciences\\
620 Cherry St NW, Room G10\\
Atlanta, GA 30312\\
cecarr@gatech.edu
\thanks{This work has been submitted to the IEEE for possible publication. Copyright may be transferred without notice, after which this version may no longer be accessible.}
}

\maketitle

\thispagestyle{plain}
\pagestyle{plain}

\maketitle

\thispagestyle{plain}
\pagestyle{plain}

\begin{abstract}
Small, low-cost instruments enable new and exciting mission opportunities, yet their constrained volume and limited budgets make them especially susceptible to suffering anomalies during flight. Radiation effects, as well as sensor or actuator failure, can all pose a serious threat to the continued collection of scientific data as well as cause the partial or complete loss of a mission’s science payload. Onboard anomaly detection could allow instruments to recover from such events, but its ad-hoc development typically falls outside the mission’s timeline or monetary constraints. Here we describe a compact solution for the implementation of onboard anomaly detection meant for space science missions. The device is designed to be interoperable with a broad range of instruments, utilizing easily accessible power and logic signals to monitor the state of peripherals and actuators without disrupting their functionality. By leveraging a commercially-available microcontroller with a radiation-hardened alternative package, the device can be inexpensively sourced and assembled with minimal work, enabling instrument characterization on an expedited timeline. The system can then be exchanged for a radiation-hardened version, ensuring the replicability of observed anomalies in a laboratory environment during instrument operations. We also present currently implemented anomaly detection algorithms, which enable the system to detect anomalies in instruments with varying failure modes and allow mission designers to choose which detection approach best fits the specific needs of their instrument. Finally, we showcase an example application of this system in the detection of anomalies during the operation of a lysis motor designed for use in biological space instruments. The inclusion of the described anomaly detection system into new or existing instruments can effectively lower the risks associated with in-flight anomalies, improving their reliability while causing minimal impact on their development timeline or system complexity. This newfound capability can be leveraged to improve the resilience of existing science missions or to enable new missions to harsher space environments where anomaly detection is a requirement for successful instrument operation.
\end{abstract}

\tableofcontents

\section{Introduction}

Small instruments are rapidly expanding our ability to achieve science objectives with reduced budgets and constrained timelines. This creates exciting opportunities to pursue new mission concepts that were previously not achievable, where mass, volume, or cost constraints limit the use of larger instruments. Missions concepts similar to the Europa Lander \cite{noauthor_europa_nodate}, the Enceladus Orbilander Mission \cite{mackenzie_enceladus_2021}, or the Vertical Entry Robot for Navigating Europa (VERNE) \cite{bryson_vertical_nodate} rely on small instruments to achieve their mission goals. Furthermore, instruments that are not yet implemented in specific mission architectures, like the ELIE \cite{carr_solid-state_2022} or SETG \cite{lui_setg_2011} instruments, have the potential to improve or enable new measurements and observations to answer still unanswered questions in planetary science. Small science missions are an integral part of current and future strategies to advance in-situ investigation of planetary or small-body environments, as they provide a valuable combination of launch cadence, complexity, and risk that enables them to respond to ongoing scientific development \cite{committee_on_the_planetary_science_and_astrobiology_decadal_survey_origins_2022}. 

However, these instruments pose unique implementation challenges for mission designers, one of which is reliable anomaly detection that can adapt to the requirements of these instruments. During the course of a mission, instruments will encounter anomalies that can threaten their successful operation, the collection of data, and the achievement of scientific goals. An anomaly is an external or internal event that changes the state or configuration of an instrument to one that can cause temporary or permanent damage to the instrument, or under which the instrument was not designed to operate. Anomalies pose a risk to all spacecraft, but their impact on small missions can be severely detrimental: Between the years 2000 and 2016, 35 percent of small missions failed to meet their objectives \cite{jacklin_small-satellite_2019}. Detecting an anomaly is vital to allow the instrument or spacecraft to respond to it, mitigating its disruption to ongoing operations and ensuring the safety and continuity of the mission. \\

\begin{figure}[h!]
    \centering
    \includegraphics[width=3.25in]{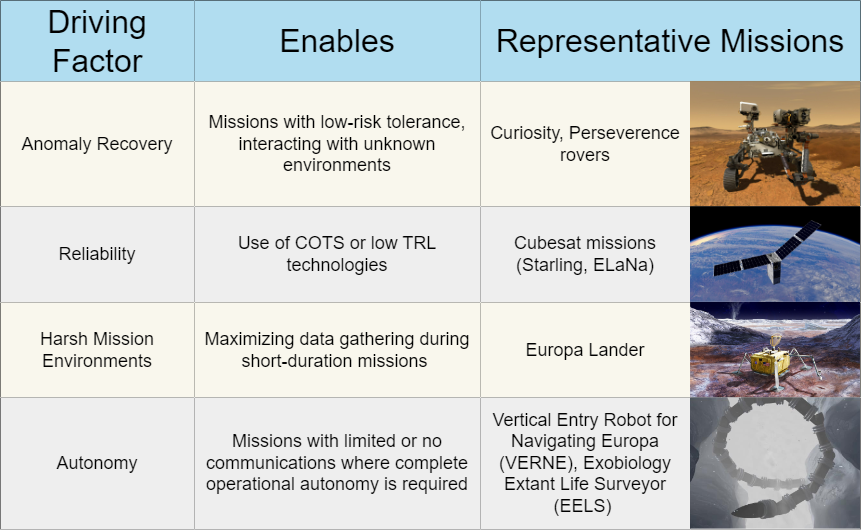}
    \caption{Mapping of driving factors, capabilities, and representative missions enabled by anomaly detection. Mission renders sourced from \protect\cite{mars_nasa_govMars2020Perseverance}, \protect\cite{noauthor_nasas_nodate},  \protect\cite{noauthor_europa_nodate} and \protect\cite{noauthor_exobiology_nodate}}
    \label{fig:anomalyDetectionDrivingFactors}
\end{figure}

The need for anomaly detection can be driven by multiple mission requirements. Recovery from anomalies into a functional or safe state can be a requirement for missions where the occurrence of an anomaly threatens the continuation of the mission or the achievement of its mission goals unless corrective action is taken. It can also be a requirement for instruments where the medium to sample is a-priori unknown. Detecting an anomaly in the instrument's operation can indicate the presence of an unfamiliar sample medium, allowing mission controllers to halt operations and conduct further studies before the instrument proceeds with sample analysis, thus mitigating the risk of damage to the instrument. A requirement for increased reliability can also drive the need for anomaly detection capabilities. Resource restrictions drive instrument designers to use commercial-of-the-shelf (COTS) or low technology readiness level (TRL) technologies in instrument designs which increases the risk of anomalies negatively impacting instrument operations. This can also affect the overall risk of the instrument and make it unsuitable for missions with low-risk tolerances. Anomaly detection can increase the reliability of these instruments and lower their overall risk. Furthermore, missions destined for harsh environments where a short mission duration is expected can leverage anomaly detection to maximize the total amount of data gathered by detecting and reporting anomalies that might affect the integrity of the data. Finally, missions requiring complete autonomy of the spacecraft and instrument due to limited or a complete lack of communication for extended periods of time must be able to react to anomalies that might threaten the successful completion of the mission objectives, which is enabled by anomaly detection. These driving factors for anomaly detection, as well as examples of missions enabled by anomaly detection capabilities, are summarized in Figure \ref{fig:anomalyDetectionDrivingFactors}.

Anomaly sources that impact an instrument can be caused by the environment the spacecraft operates in, like radiation-related latch-up events or bit flips, or by the failure or misconfiguration of a system or subsystem that is part of the spacecraft itself \cite{galvan_satellite_2014}. Finally, operator error and faulty hardware design threaten the correct function of the instrument. Due to this variability in anomaly sources, there is no single method that can detect all possible anomalies an instrument might experience during the duration of its mission. \\

Current anomaly detection solutions rely on the ad-hoc implementation of both custom hardware and software on the specific system to be protected from a subset of possible anomalies. This requires an investment of both resources and time for the development of this functionality, which is sometimes infeasible for small instruments due to their limited budgets and restricted timelines. Furthermore, identifiable results of anomalies are dependent on the specific implementation of actuators and peripherals that compose the instrument, and as such customized anomaly detection solutions are needed for each instrument or instrument implementation. This need for a customized detection solution clashes with the agile timelines expected from small instrument developers. This tight implementation schedule might also prohibit the inclusion of an independent anomaly detection system, opting instead for the implementation of anomaly detection as part of the instrument software or hardware system directly. Although this implementation approach might be preferable over the complete lack of anomaly detection, it does not mitigate the risk of a fault in the instrument rendering all or part of the anomaly detection measures ineffective. Furthermore, the increased complexity of implementing anomaly detection might not be manageable for teams focused on the development of an instrument. It might also unnecessarily complicate the software and hardware design of an instrument, which itself poses a risk of faults. Current anomaly detection solutions thus fail to meet the requirements of small instruments.

Mitigating the risks of anomalies impacting small instruments would require an anomaly detection solution that is easy to integrate with a multitude of different instruments, eliminating the need for the development of unique anomaly detection solutions for each spacecraft or instrument implementation. This would reduce the time and resource commitment required by teams to implement anomaly detection, making it more accessible. To achieve this goal, the solution must also have a clear and defined interface to communicate with the rest of the instrument and be compact and inexpensive. It should also be easy to use by instrument designers and ground operators to reduce the chance of anomalies caused by improper implementation. Finally, this generic anomaly detection solution that provides this capability to multiple instruments must meet the diverging requirements found during instrument development in the laboratory and during flight while minimizing complexity and risk to replicability.

In the laboratory environment, the anomaly detection system must be easy to procure and inexpensive to enable fast development and simplify the instrument characterization process. Usability by operators also becomes a concern during laboratory testing, as the output of the anomaly detection system should be easily interpretable for rapid development. During flight, the need for systems to be radiation tolerant or radiation hardened becomes a top priority, especially for missions with objectives past low Earth orbit. These competing requirements would normally force teams to devise two separate systems: one to characterize their instrument in the laboratory and a separate system to provide anomaly detection during flight. This increases the complexity throughout the instrument's life cycle, but more importantly, it poses a risk to the replicability of unexpected fault states experienced during flight: As the flight and laboratory systems are fundamentally different, recreating an anomalous state in the laboratory using information from flight requires understanding how to translate the flight system's output into the expected input for the laboratory system.

The proposed anomaly detection solution allows instrument designers to use the same solution to characterize expected anomalies and other instrument parameters prior to hardware handoff and to detect anomalies during flight using the same instrument subsystem.  During assembly and laboratory instrument testing, anomaly detection can be used to characterize the expected operational states of the instrument and profile known or expected failure states that might result from possible hardware design limitations or flaws. Failures resulting from random events like single event effects (latch-ups or bit flips) can also be emulated in the laboratory to determine the resiliency of the instrument to such faults. During the instrument's use in a mission, the anomaly detection system can be used to detect anomalous states during instrument operation, which allows the instrument to respond to failures and recover from states that might pose a risk to its safety and continued functionality. An anomaly detection solution can also provide instrument monitoring capabilities useful for instrument operation independent of the instrument itself.\\

\section{Methods}


\subsection{System Architecture}

\begin{figure}[h!]\label{OneColumn}
\centering
\includegraphics[width=3.25in]{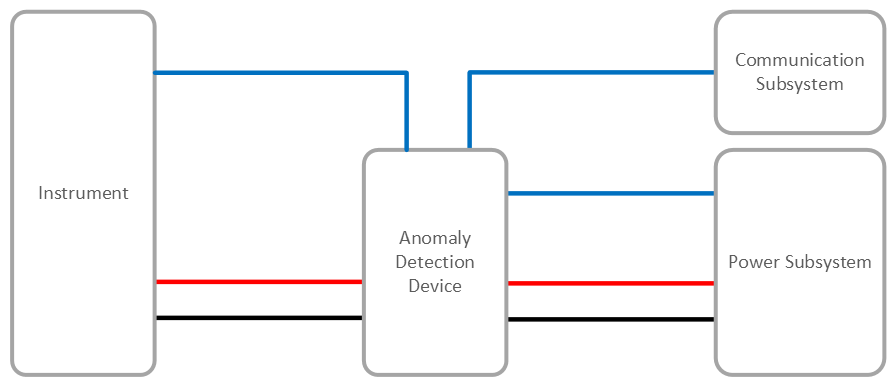}\\
\caption{\textbf{High-level anomaly detection system architecture, showing the connection between the anomaly detection device and other subsystems.}}
\label{fig:anomalyDetectionSystemDiagram}
\end{figure}

The proposed anomaly detection device follows a decoupled architecture, where the anomaly detection function is performed not as part of the instrument or power systems but as an independent device with separate hardware and software design. This allows the device to be used for different instruments without the need to have a specific hardware implementation for each one. Depending on the role anomaly detection is meant to play in the mission, the device can be used as the starting node of a more expansive monitoring and reliability system that oversees the functionality of the instrument. It can also be connected to the instrument or other systems on the spacecraft for closed-loop control of instrument modes or to a communication subsystem for off-board instrument monitoring, as shown in the system diagram presented on Figure \ref{fig:anomalyDetectionSystemDiagram}. 

\begin{figure}[h!]\label{OneColumn}
\centering
\includegraphics[width=3.25in]{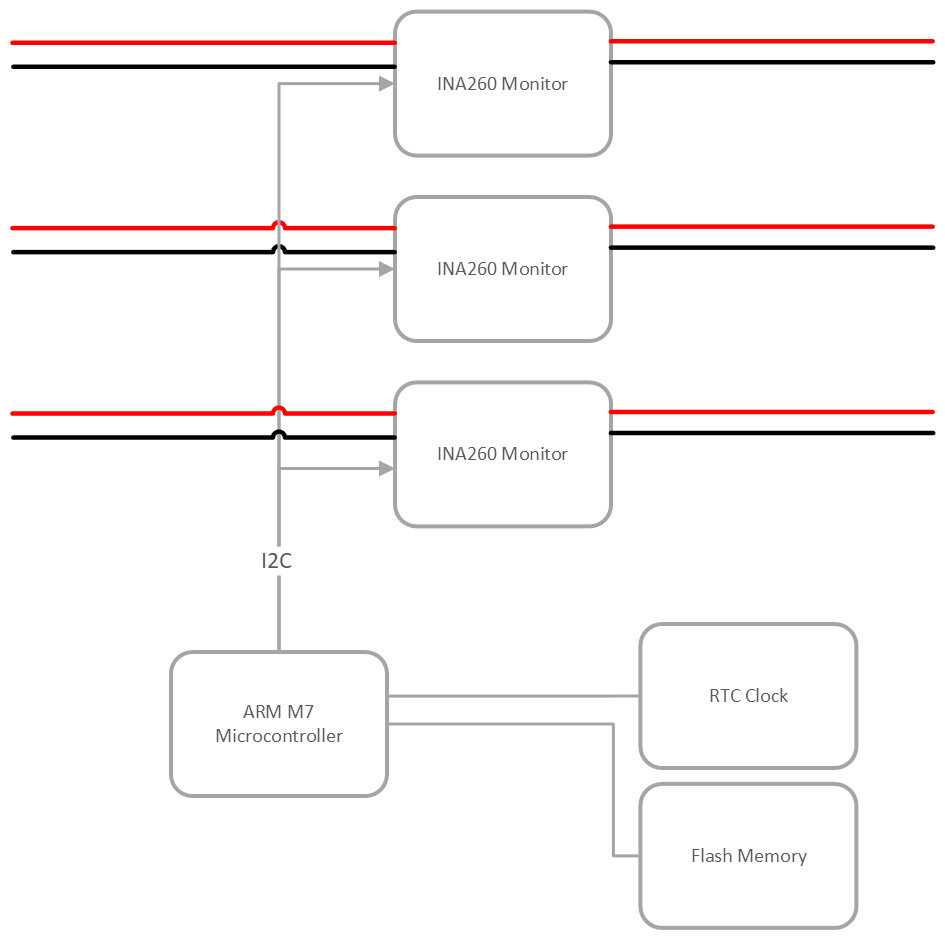}\\
\caption{\textbf{Anomaly detection device block diagram when connected to multiple INA260 sensors.}}
\label{fig:multipleINA}
\end{figure}

The device detects anomalies in the connected instrument by measuring its power consumption over time. This measurement is taken using an INA260 Precision Digital Current and Power Monitor, which is managed by an ARM M0 Cortex microcontroller. The data is processed by the microcontroller and time-tagged through the use of a real-time clock (RTC), after which it is stored in external flash memory and/or sent to a communication system through I2C or SPI. Multiple INA260 sensors can be connected to the same device as shown in Figure \ref{fig:multipleINA} if the instrument uses more than one power rail to operate. The detection of anomalies through this data is dependent on the specific implementation of the anomaly detection device and the type of anomaly to be detected. These will be discussed further in the System Use and Anomaly Detection Algorithms sections. The detection and monitoring of anomalies are performed by the microcontroller which can then alert the instrument or supporting systems through the aforementioned communication protocols.

\begin{figure}\label{OneColumn}
\centering
\includegraphics[width=3.25in]{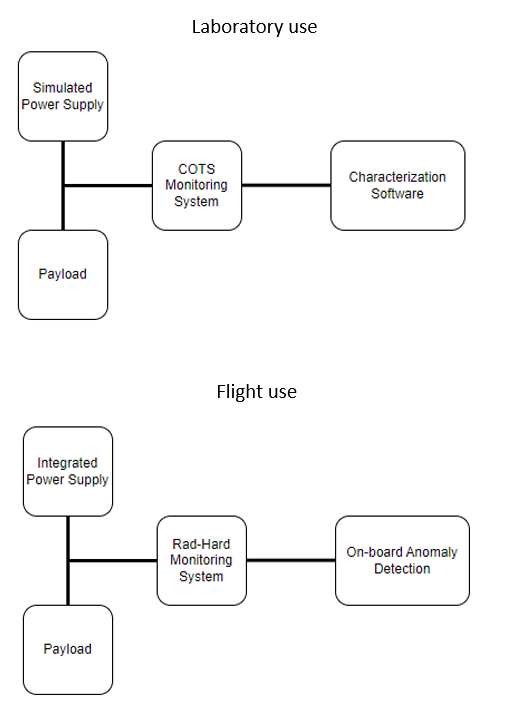}\\
\caption{\textbf{System diagrams for laboratory and flight use of the device, showing the connections to the instrument and supporting systems.}}
\label{fig:labAndFlightDiagram}
\end{figure}

The device can be used through two different hardware implementations, depending on its intended use: a laboratory implementation, which uses inexpensive and readily available COTS parts and microcontrollers; and a flight version which uses a radiation-hardened microcontroller to enable deep space missions. The device is implemented on two distinct pieces of hardware to best meet the design constraints of the laboratory and flight environments while minimizing the risk of fault replicability associated with utilizing different hardware. Figure \ref{fig:labAndFlightDiagram} shows the connections between the device, the instrument, and its power system under the two different hardware implementations. The current laboratory implementation fits within a $6.4 \times 2.4 \times 1.9$ inch volume, weights $20$ grams, and consumes an average of $25.2 mW$ during active use.

\subsection{Integration with an instrument}
As the device relies on the INA260 sensor to monitor power, the device must be connected to the power supply of the instrument to monitor the instrument's state and detect anomalies. This can be done on either the high or low side of power, depending on other mission requirements, and will not interfere with any non-signal carrying power supply. The device can also be connected to the instrument's peripherals, actuators, or sensors which designers deem to be at the highest risk of experiencing a fault. This provides even more granularity in the detected anomalous states. A total of 16 INA260 sensors can be connected in parallel to a single monitoring device, enabling a wide survey of all potential sources of faults present in an instrument. 

\subsection{System utilization}

The device is first connected to the instrument during final integration in the laboratory, after which the instrument can be put through functional testing. As part of this testing, expected modes of operation and known possible faults can be profiled, which allows the anomaly detection system to be configured for the specific needs of the instrument at hand. Expected modes of operation to be profiled might include the power draw from the instrument during standby, startup, or expected conditions during data collection. Faults to be tested can include possible actuator or active component failures, or mission subsystem failures that might threaten the nominal state of the instrument. During this process, the anomaly detection device is connected to a computer running a profiling and configuration application which collects data from the device in real time and allows for the tagging and archiving of instrument monitoring data. This data can then be used to produce a configuration for the anomaly detection device that can be used during flight. \\


Once the instrument has been characterized and possible fault modes have been used to configure the anomaly detection device, the corresponding configuration can be loaded onto the flight version of the anomaly detection device and subsequently integrated with the instrument before hardware hand-off. The application code used by both microcontrollers is reused, with only the Hardware Abstraction Layer (HAL) and middleware required to drive the anomaly detection peripherals (INA260 sensors, RTC, and flash memory) regenerated. Although this inherently poses a risk to the laboratory replicability of errors experienced in flight, sacrificing the use of a one-to-one replica of the instrument in exchange for lower costs, ease of procurement, and fast integration with the instrument might be a positive trade off for small instruments with lax code validation and verification requirements. \\

\subsection{Anomaly detection algorithms}

The capability of the anomaly detection device relies on its detection algorithms. There is no singular ideal anomaly detection algorithm that can detect anomalies in all possible instruments. As such multiple approaches have been implemented to allow instrument designers to tailor the algorithms to their specific use cases. The algorithms rely on the voltage and current measurements to identify anomalous states of the instrument or instrument subsystems. The currently implemented algorithms include Out-of-limits and AutoEncoder-based anomaly detection. These were chosen as they lie on opposite sides of the available algorithm spectrum in terms of complexity, explainability, and ease of implementation, showcasing the flexible capabilities of the anomaly detection device. 

\subsubsection{Out-of-limits anomaly detection}

\begin{figure}[h!]\label{OneColumn}
\centering
\includegraphics[width=3.25in]{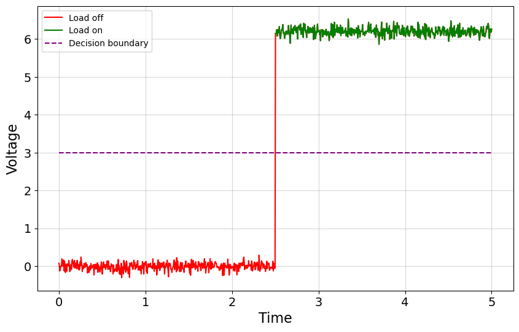}\\
\caption{\textbf{Example of out-of-limits anomaly detection on a power rail load off to load on transition.}}
\label{fig:outOfLimitsExample}
\end{figure}

For simple monitoring purposes or for anomalies that directly translate into overvoltage, overcurrent, undervoltage, or undercurrent of a system, an out-of-limits anomaly detection schema might be the most explainable and effective anomaly detection method. This involves quantifying the expected maximum and minimum values for a given voltage and current signal connected to the anomaly detection device. Once these limits have been determined, the anomaly detection device will flag an anomaly if the measured voltage and current exceed the defined limits. Figure \ref{fig:outOfLimitsExample} shows an example of the out-of-limits anomaly detection schema detecting the delivery of power to a sample load after it is connected to a power rail being monitored by the device. This can be implemented on the device's microcontroller or as an alarm directly commanded by the INA260 sensor through an I/O signal. This second option minimizes the time taken by the system to detect and report an anomaly and can be especially useful if action must be taken by the instrument in a short time to correct the fault. 

Out-of-limits anomaly detection is easy to implement and easily explainable, but it lacks the ability to detect anomalies that primarily manifest in the time domain and not in the magnitude of the signal. Furthermore, it requires the boundaries of the signal to be known before the anomaly happens, which is not always feasible for more complex anomalies. 

\subsubsection{Autoencoder-based anomaly detection}

\begin{figure}[h!]\label{OneColumn}
\centering
\includegraphics[width=2.75in]{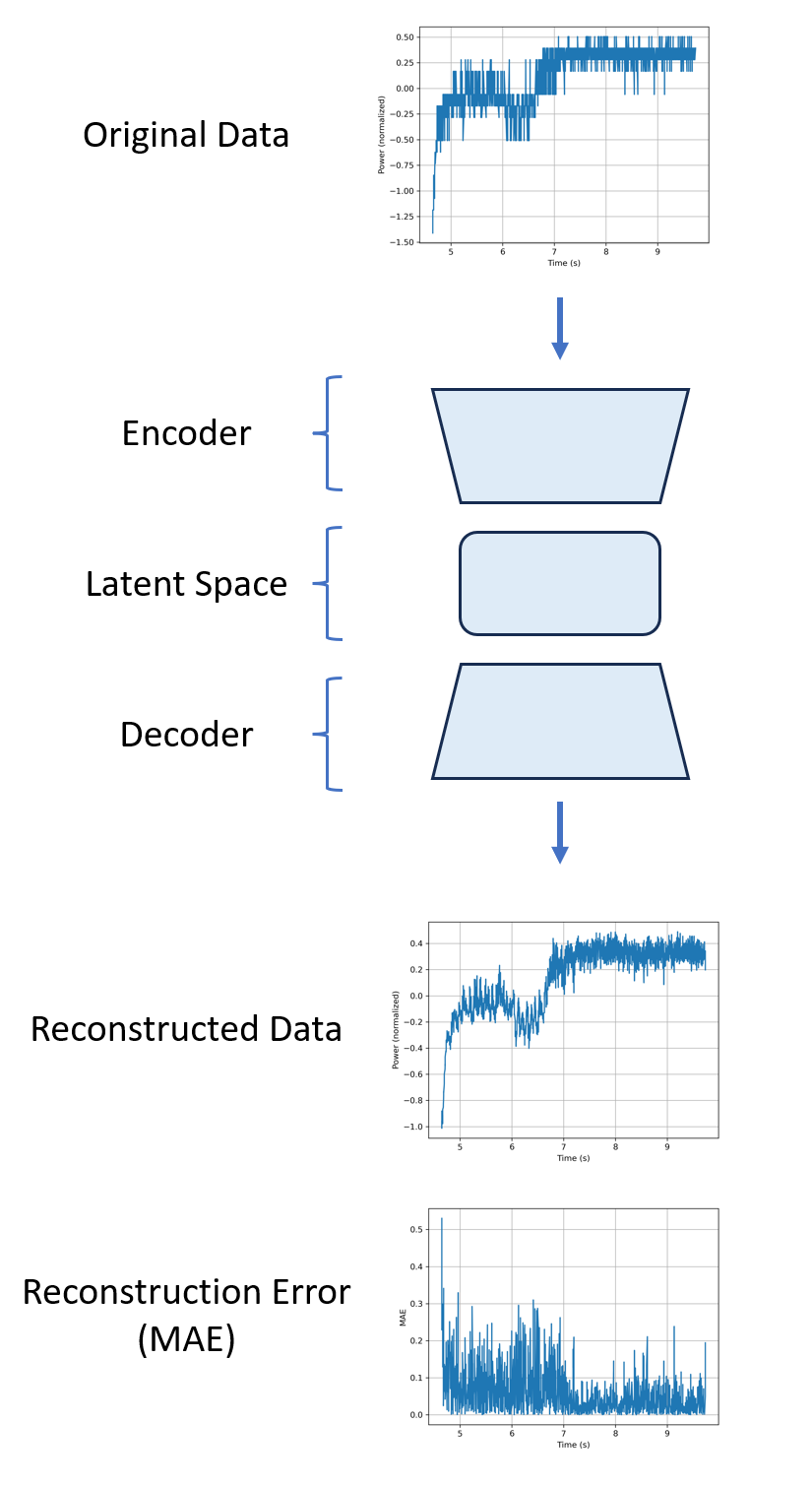}\\
\caption{\textbf{Sample autoencoder architecture, showing example data being reconstructed by the network.}}
\label{fig:exampleAuto}
\end{figure}

Autoencoders are a type of neural network whose aim is to reconstruct the input signal with as little error as possible. An autoencoder is formed by a converging section that tapers down to a latent space of reduced dimensionality. This latent space is then followed by a diverging section where layers get progressively wider until they reach the same layer width as the input. During training, the network learns to compress the input data down to the latent space, and then reconstruct it back to its original form. This architecture is presented in Figure \ref{fig:exampleAuto}.

\begin{figure}[h!]\label{OneColumn}
\centering
\includegraphics[width=3.35in]{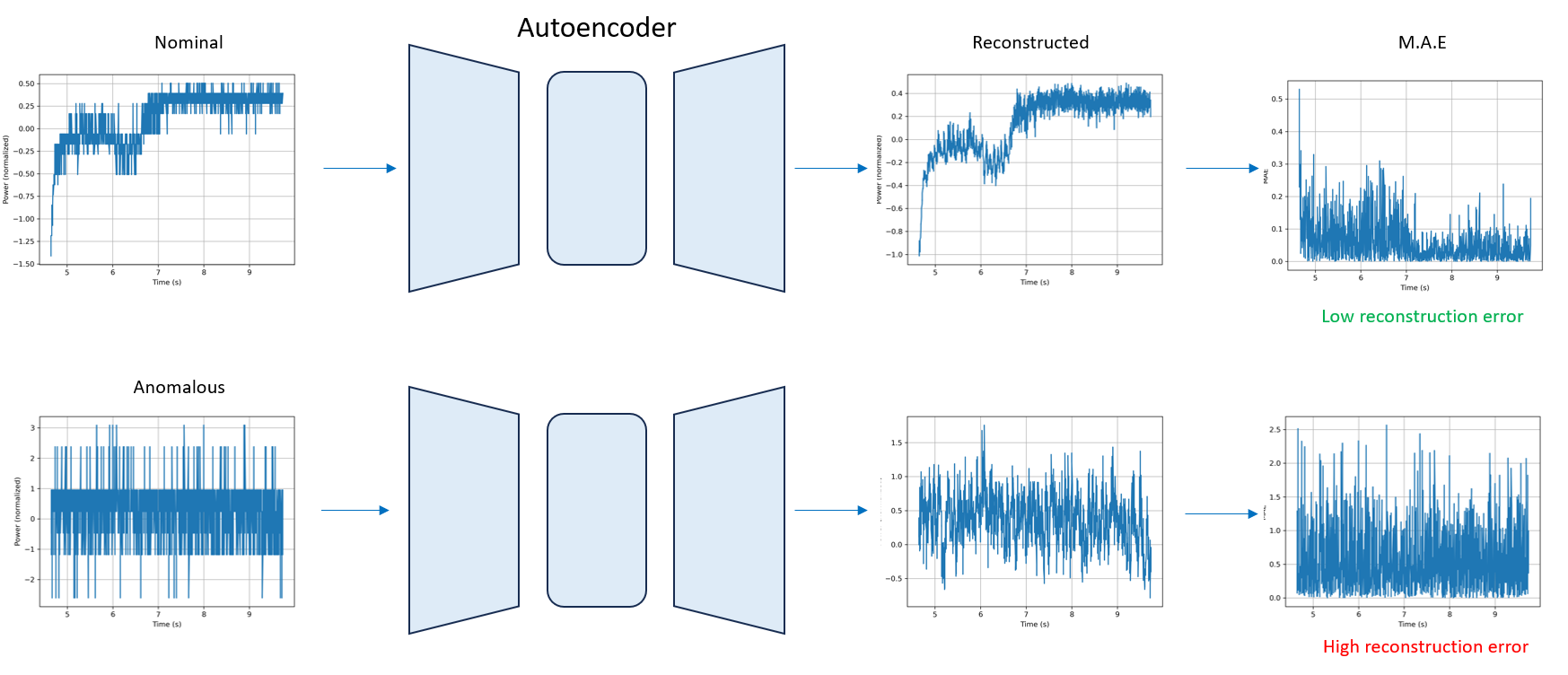}\\
\caption{\textbf{Example of autoencoder network applied to anomaly detection. The reconstruction MAE is used to determine if input data is anomalous.}}
\label{fig:anomalyAuto}
\end{figure}

Autoencoders have proven to be valuable tools for anomaly detection, as their ability to compress and reconstruct new data down to the dimensionality of the latent space is highly dependent on the characteristics of the training data \cite{thill_temporal_2021}. The reconstruction error of the autoencoder can then be used to measure the similarity of new inputs to the training data. Anomalous data that does not match the characteristics of the training data will produce a high reconstruction error, which flags the data as anomalous. An example of this workflow is shown in Figure \ref{fig:anomalyAuto}.

Autoencoders have the advantage of being able to detect anomalies not characterized during instrument testing, as any data that does not fit the training data will be flagged as an anomaly. Multiple autoencoders can be trained on different data sets to identify anomalies in instruments with multiple nominal states, providing instrument designers with a flexible anomaly detection solution that can accommodate multiple operation modes.\\

However, autoencoders suffer from a lack of explainability as their output cannot be easily interpreted to be meaningful or useful beyond characterizing an anomaly. They also require training data to be collected. The heavy computational cost of training a neural network is a limiting factor for the complexity of the system in most microcontroller systems. These issues are mitigated in our implementation of the autoencoder by utilizing dense layers with no convolution and performing the training during the laboratory procedures on a separate device and not on the microcontroller itself. The trained weights are then loaded onto the anomaly detection device and used to compute the reconstruction error of newly measured data, which indicates whether the data is nominal or anomalous.

\section{Results}

Here, we showcase the application of the anomaly detection device previously presented for the detection of anomalies in the operation of a lysis motor. This component is being used as part of the sample processing stage of the ELIE instrument \cite{carr_solid-state_2022} and is a common component in other life detection instrumentation under development. In principle, this lysis motor could be replaced with any given load that exhibits changing power draw as an unknown function of state. This includes most actuators, peripherals, or subsystems with electrical components present in instruments or spacecraft. 

\subsection{Background}
Lysis is the process through which the membranes of cells are broken down, which makes materials only present inside the cell available for further analysis, such as DNA, RNA, proteins, or organelles. This process can be induced mechanically through the use of a lysis motor, which combines a small electrical brushed motor with a lysis chamber containing grinding beads through which a solution with the cells of interest is pumped. The motor stirs the grinding beads, which break down the cell walls and produce a lysate (solution containing the products of lysis) which can then be used for further analysis. The motor relies on the constant flow of fluid through the lysis chamber to avoid overheating during operation but does not pump the fluid itself and instead relies on a pumping system to continue feeding fluid through the lysis chamber.\\

A failure of the pumping system which starves the motor of fluid for a continued period of time will cause it to overheat, damaging it and preventing its further use. Furthermore, a latch-up event can maintain the motor running past the period through which the pumping system is providing it with fluid, causing it to overheat still. As fluid systems take up significant amounts of space and increase the complexity of an instrument, being able to reliably operate a single motor and fluidic line is critical. These two possible failure modes were used to validate the anomaly detection algorithms currently implemented on the anomaly detection device while showcasing its setup, workflow, and features.

\subsection{Experimental setup}
\begin{figure}[h!]\label{OneColumn}
\centering
\includegraphics[width=3.25in]{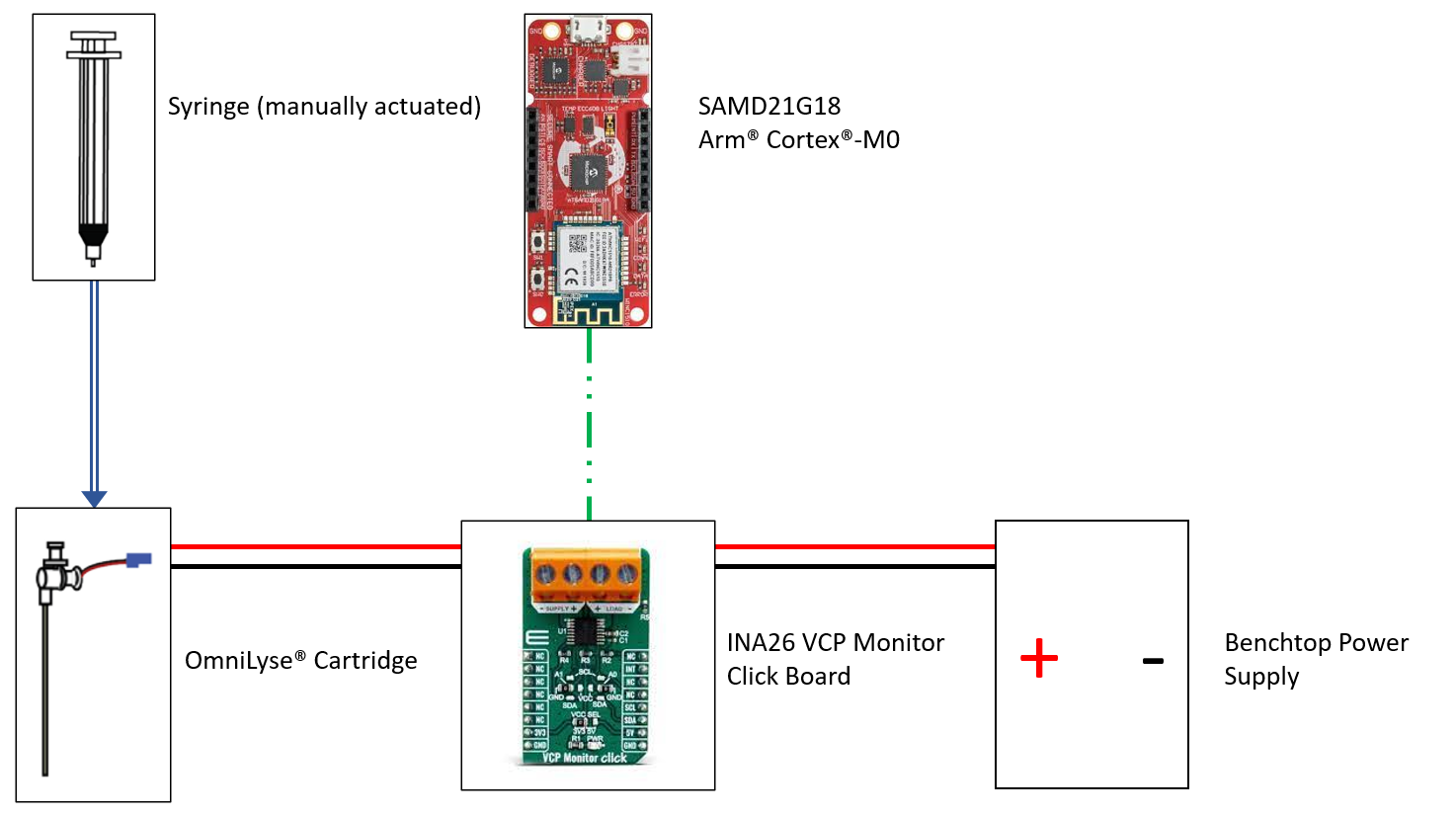}\\
\caption{\textbf{Lysis motor experimental setup used to collect training data.}}
\label{fig:expSetup}
\end{figure}

The anomaly detection device was first connected between an OmniLyse® lysis motor and a power source, a benchtop power supply. A syringe connected to the motor was used to control the flow of fluid through the lysis chamber, as shown in the setup diagram presented by Figure \ref{fig:expSetup}. Instead of a cell solution, water was used as a cooling fluid. The microcontroller used for this setup was a Microchip SAMD21G18 Arm® Cortex®-M0 implemented on an IOT development board. The INA260 sensor was connected to the microcontroller through a Click development board and the power supply for the motor was routed through the sensor.

The motor was first turned on and off repeatedly to capture data on the startup and shutdown behavior of the lysis motor. The motor was turned on and off roughly every 10 seconds by switching the power supply on and off, for a total duration of 50 seconds. This data was then used to determine a power threshold for when the motor was operating and when it was turned off. The state of the motor can be found utilizing the anomaly detection device and its out-of-bounds algorithm configured with the measured threshold.\\

The motor was then turned on constantly and run both with fluid constantly flowing through the lysis chamber (in wet conditions) and with no fluid flowing through (dry conditions). Readings for wet and dry conditions were captured in 60-second intervals, varying the voltage between 6V and 12V. An autoencoder network was trained on a subset of the wet conditions data. The rest of the wet and the dry data were then used as inputs to the network and the mean square error of the resulting data reconstruction was recorded.

\subsection{Outcomes} 

\begin{figure}[h!]\label{OneColumn}
\centering
\includegraphics[width=3.25in]{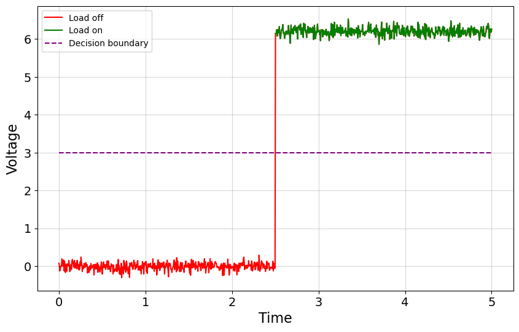}\\
\caption{\textbf{Detection of motor operation through out-of-limits anomaly detection.}}
\label{fig:motorOOL}
\end{figure}

Data from the first trial was processed using the out-of-bounds algorithm with an on/off threshold value of 3V. The algorithm was able to correctly determine the state of the motor, showing the anomaly detection device can correctly determine the running state of the motor from power consumption data, as shown in Figure \ref{fig:motorOOL}. This capability is of special interest for anomalies that cause the commanded and actual state of a peripheral to not correspond to each other, such as latch-up events, as it provides a direct measurement of the running state of the peripheral independent of the commanded value. 

\begin{figure}[h!]\label{OneColumn}
\centering
\includegraphics[width=3.25in]{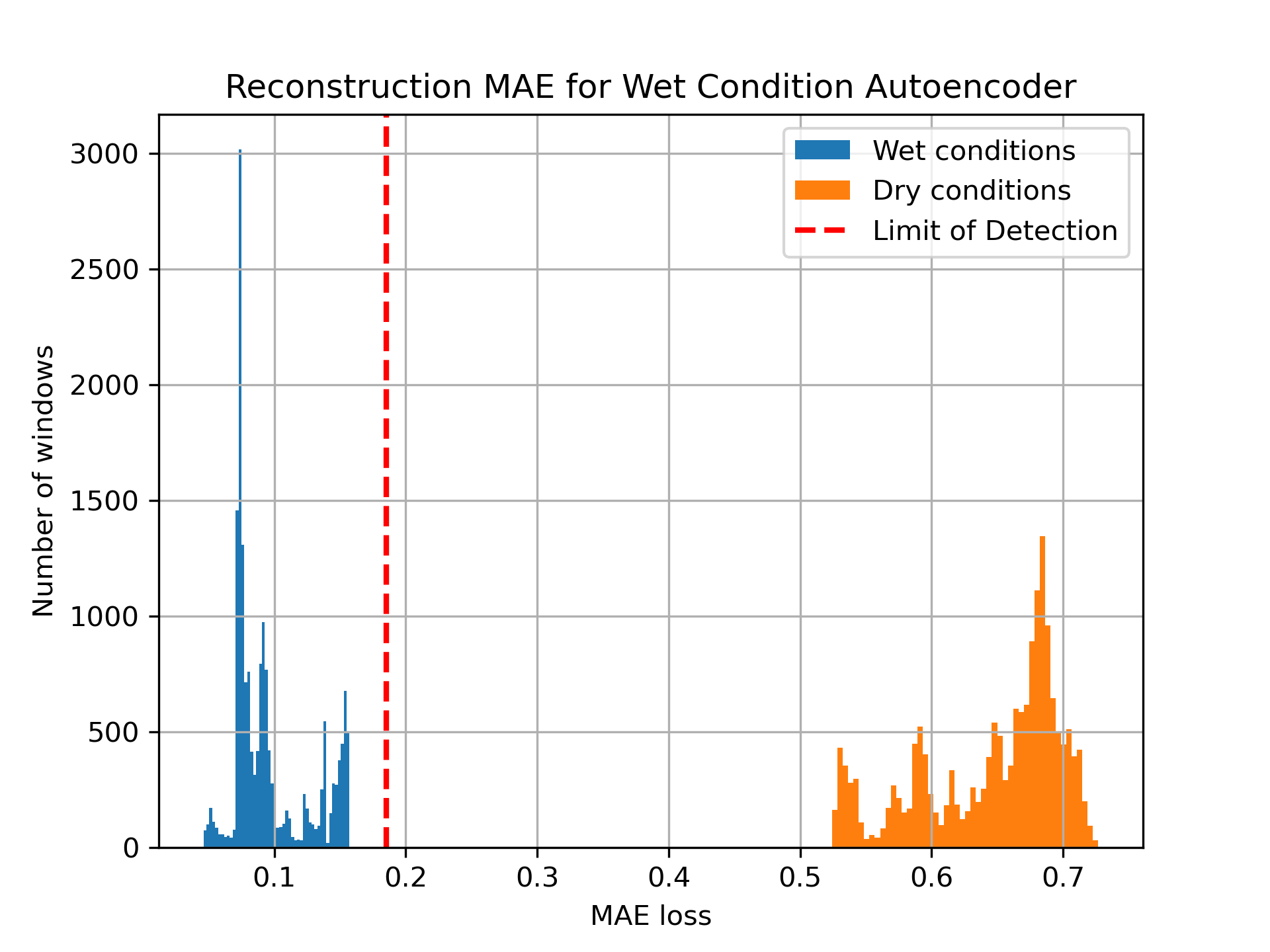}\\
\caption{\textbf{Detection of dry or wet conditions through autoencoder anomaly detection.}}
\label{fig:motorautoencoder}
\end{figure}

Data from the second trial was first normalized and processed into rolling windows with a 5-second duration. The rolling windows resulting from the wet running conditions test were used to train an autoencoder network, which effectively reconstructed the rolling windows after training, with a mean absolute error of $0.1 mW$ across all measured voltages under wet running conditions. This autoencoder network was then used to reconstruct dry running condition data, yielding reconstructions with an increased MAE (mean absolute error) of $0.64 mW$ across all measured voltages. The mean absolute errors of all sampled windows were used to capture the distribution of the autoencoder reconstruction loss as a function of the operating conditions of the motor during data capture. The distribution of the wet windows was used to compute a limit of detection (LOD) for anomalies in the motor operating conditions, which was taken to be the mean plus three times the standard deviation of the wet conditions reconstruction MAE. The distributions of the MAE during dry and and wet running, as well as the LOD are presented in Figure \ref{fig:motorautoencoder}. The reconstruction loss of dry running condition data shows the effectiveness of the autoencoder at detecting anomalous motor operation conditions, as measurements taken during dry running conditions correctly lie past the LOD.

\section{Conclusion}

Anomaly detection is a critical component of any mission, as it protects instruments from irreparable damage caused by unexpected events that will inevitably occur during their lifetime. The presented anomaly detection device provides small instrument designers and integrators with an effective, inexpensive, and compact solution to add this capability to their instruments without investing large amounts of work implementing an ad-hoc solution for their instrument. The ability to use the same system in a laboratory setting as well as during flight gives users an edge to lower costs and reduce time to develop and deliver instruments, as well as ensuring anomalies observed in flight can be replicated. The proposed system was demonstrated for use with an instrument component that is analogous to actuators or peripherals readily present in other instruments and spacecraft, showcasing the ease of use and adaptability of the device. Continued development of this anomaly detection solution and its implementation in instruments through the use of radiation-hardened hardware can enable new missions where anomaly recovery capabilities and heightened reliability are needed to access harsh mission environments autonomously, to further advance in-situ investigation of open questions in planetary science.


\bibliographystyle{IEEEtran}
\newcommand\BIBentryALTinterwordstretchfactor{1}
\bibliography{Bibliography}





\thebiography
\begin{biographywithpic}
{Alfonso Lagares de Toledo}{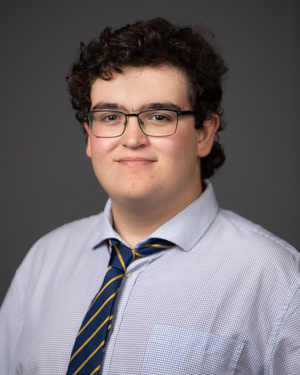}
 is an undergraduate
researcher at Georgia Tech pursuing a
B.S. degree in Aerospace Engineering. He is interested in leveraging
system design to improve the way avionics are integrated into space missions.
\vspace{30px}
\end{biographywithpic} 
\begin{biographywithpic}
{Christopher E. Carr}{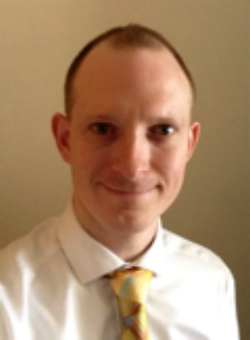} received his B.S. degree in Aero/Astro and Electrical Engineering in 1999, his Masters degree in Aero/Astro in 2001, and his Sc.D. degree in Medical Physics in 2005, all from MIT. He is an Assistant Professor in the Daniel Guggenheim School of Aerospace Engineering and the School of Earth and Atmospheric Sciences at the Georgia Institute of Technology. He is broadly interested in searching for and expanding the presence of life beyond Earth.
\end{biographywithpic}

\end{document}